\documentclass[%
 reprint,
 amsmath,amssymb,
 aps,
]{revtex4-2}

\usepackage{graphicx}
\usepackage{dcolumn}
\usepackage{bm}
\newtheorem{hypothesis}{Hypothesis}

\begin{document}

\preprint{APS/123-QED}

\title{Interpretation of the superposition principle and locality loophole in Bell experiments}

\author{Sheng Feng}
 \email{fengsf2a@hust.edu.cn}
\affiliation{%
School of Electrical and Electronic Information Engineering, Hubei Polytechnic University, Huangshi, Hubei 435003, P.R. China\\
}%




\date{\today}

\begin{abstract}
A connection is revealed between the superposition principle and locality. A self consistent interpretation of the superposition principle is put forth, from which it is shown that quantum mechanics may be a local statistical theory. Then it is shown how Bell experiments can be satisfactorily explained by assuming local nature for entangled particles, i.e., the violation of Bell inequality cannot distinguish between locality and nonlocality, which is referred to as locality loophole. Moreover, existing experimental results are presented indicating locality in quantum mechanics and new experiments are proposed so that the locality loophole may be closed. 
\end{abstract}

\maketitle


\section{Introduction}
As an essential property of quantum mechanics, nonlocality has come into being as a consequence of the far-reaching debate between Bohr and Einstein on the completeness of the quantum mechanical description of reality \cite{epr,Bohr1935}. After the seminal inequality given by Bell for an upper bound on the strength of some correlations exhibited by local realist theories \cite{Bell1964}, Bell experiments violating this inequality or its variants have been considered as the standard platform for the test of nonlocality  \cite{Freedman1972,Aspect1982,Greenberger1990,Giustina2015,Shalm2015,Hensen2015}. However, the derivation of Bell inequality in fact assumed both locality and predetermination (reality) at the same time \cite{Bell1964}. So, strictly speaking, violation of Bell inequality can reject at least one of the two assumptions, but not necessarily both. For instance, it is in principle possible that Bell experiments reject only the reality  assumption with the locality assumption survived. Therefore, in view of the fundamental importance of the nonlocality concept in quantum mechanics \cite{Bohm1952,Pitowsky1982,Walborn2011,Cavalcanti2011,Christensen2013,Hirsch2013,Erven2014,Brunner2014,Popescu2014}, it would be of great interest to persue different lines of inquiry into the topic such that the concept may be more consolidated.

To that aim, this work first dives into the self consistency in the interpretation of the superposition principle, leading to several hypothesises that reveal a deep connection between the superposition principle and locality. From this connection it follows that quantum mechanics could be a local statistical theory. Using the proposed hypothesises, it is further shown that violation of Bell inequality can be well explained by assuming local property for entangled particles. This confirms the aforementioned conjecture that Bell experiments do not reject the locality assumption and hence opens a fundamental loophole of locality. In addition, existing experimental results will be presented that seem to indicate locality in quantum entanglement, which strongly urges both theoretical and experimental research efforts to close the loophole. For this purpose, two types of experiment are proposed.

\section{Interpretation of the superposition principle and locality}
To begin with, let consider a particle in a quantum state $|\psi\rangle$ as a superposition of other states,
\begin{equation}
|\psi\rangle=\sum_{j=1}^m a_j |\varphi_j\rangle\ ,
\label{eq:superposition}
\end{equation}
wherein $m$ is a positive integer, $a_j$ are complex numbers satisfying $\sum_j|a_j|^2=1$, and $|\varphi_{j}\rangle$ represent the eigenstates of an operator $\hat{A}$ describing a measurement that may be performed on the particle. In a popular interpretation of the superposition principle, the following hypothesis is widely assumed, 
\begin{hypothesis}
A particle described by Eq. (\ref{eq:superposition}) is simultaneously in all $|\varphi_j\rangle$ states with the probability of $|a_j|^2$ for each corresponding state. 
\end{hypothesis}
Albeit this interpretation of the superposition principle may be thought to be orthodox, {\bf hypothesis 1} is unfortunately not self consistent in theory, as will be shown below. To that end, let examine a specific example of an electron in a superposed spin state,
\begin{equation}
|\phi\rangle=\left(|+_z\rangle + |-_z\rangle\right)/\sqrt{2}   \ ,
\label{eq:superspin}
\end{equation}
where $|\pm_z\rangle$ denote the spin-up and spin-down states along the z-axis, respectively, i.e., the eigenstates of $\hat{\sigma}_z$, the z-component of the Pauli operator that is a vector in parallel with the spin operator. From {\bf hypothesis 1} it follows that the electron is simultaneously in both $|\pm_z\rangle$ states with equal probability of 50\% for each state. The self inconsistency of {\bf hypothesis 1} comes from the fact that the state of the electron described in Eq. (\ref{eq:superspin}) can be equivalently re-expressed as a superposition of other spin states,
\begin{equation}
|\phi\rangle=e^{-i\pi/4} \left(|+_y\rangle + |-_y\rangle\right)/\sqrt{2}   \ ,
\label{eq:superspiny}
\end{equation}
in which $|\pm_y\rangle$ stand for the eigenstates of the y-component of the Pauli operator, $\hat{\sigma}_y$. From Eq. (\ref{eq:superspiny}) and {\bf hypothesis 1}, one must admit that the electron is simultaneously in both $|\pm_y\rangle$ states with equal probability of $|c_\pm|^2$=50\% in each state, which is obviously inconsistent with the previous statement that the electron is also simultaneously in $|\pm_z\rangle$ states with 50\% probability in each state.

To solve the issue of self inconsistency, one must note that Eqs. (\ref{eq:superspin}) and (\ref{eq:superspiny}) are equally true for the electron state in the context of {\bf hypothesis 1}. In view of this, a mathematical expression for the superposed state should take into account all the probabilities for the electron to be in the $|\pm_z\rangle$ and $|\pm_y\rangle$ states, 
\begin{equation}
|\phi\rangle=(|+_z\rangle + |-_z\rangle)/\sqrt{8}+e^{-i\pi/4}(|+_y\rangle + |-_y\rangle)/\sqrt{8}   \ .
\label{eq:superspinyz}
\end{equation}
Eq. (\ref{eq:superspinyz}) is unfortunately very odd in the sense that  $|\phi\rangle$ is not expanded in orthogonal basis and the coefficient for each term on the right hand side does not have the same physical meaning as those in Eqs. (\ref{eq:superspin}) and (\ref{eq:superspiny}). Despite of its oddness, Eq. (\ref{eq:superspinyz}) articulates that the electron is in each of the four spin states with equal probability of no more than 25\% for each state. But this is contradictory to {\bf hypothesis 1} according to which the electron is in $|\pm_z\rangle$ (or $|\pm_y\rangle$) states with 50\% probability.

The above inconsistency problem actually roots in {\bf hypothesis 1} in that it overlooks the implicit equivalence of Eqs. (\ref{eq:superspin}) and (\ref{eq:superspiny}) for the electron state; this equivalence needs to be abolished to avoid the self inconsistency in the interpretation of the superposition principle. To that end one needs to introduce some extra element such as measurement into the interpretation. Then one may arrive at a revised version of hypothesis as follows,
\begin{hypothesis}
A particle described by Eq. (\ref{eq:superposition}) is simultaneously in all possible $|\varphi_j\rangle$ states, with the corresponding probability of $|a_j|^2$ that the particle is in the $|\varphi_j\rangle$ state if an $\hat{A}$-measurement is performed on it. 
\end{hypothesis}
According to {\bf hypothesis 2}, the equivalence of Eqs. (\ref{eq:superspin}) and (\ref{eq:superspiny}) is broken when for instance a $\hat{\sigma}_z$-measurement is carried out on the electron, in which case only Eq. (\ref{eq:superspin}) holds true for predicting the probability with which the electron will be in $|\pm_z\rangle$ states after the measurement. Although progress has been made towards a self-consistent interpretation of the superposition principle, {\bf hypothesis 2} lacks a clear statement about the probability that the particle is in each $|\varphi_j\rangle$ state before the $\hat{A}$-measurement. Another potential issue is that {\bf hypothesis 2} neglects the case in which the measurement result is known to the experimentalist who performs the $\hat{A}$-measurement and hence the probability becomes 100\% that the particle is in the specific known $|\varphi_j\rangle$ state.

To clarify the above ambiguities, one must be aware of that the term ``probability" is used to quantitatively the describe the chance that a given event occurs with some uncertainty. So, for the purpose of this work, one should take into account not only the uncertainty of the measurement results but also the uncertainty of the measurement itself as long as it has not been performed. The idea may be exemplified with the electron in the studied $|\phi\rangle$ state, for which there is always a chance that a $\hat{\sigma}_z$-measurement is to be replaced by a $\hat{\sigma}_y$-measurement unless the $\hat{\sigma}_z$-measurement is actually done. This is true for even a planned measurement from which the electron cannot distinguish a unplanned one, i.e., all measurements of the same type, no matter they are planned or not, are equally possible for the particle. Therefore, $|\pm_z\rangle$ and $|\pm_y\rangle$ states are also equally possible for the electron before a spin measurement, despite that Eq. (\ref{eq:superspin}) will be preferred over Eq. (\ref{eq:superspiny}) after a $\hat{\sigma}_z$-measurement. So, one must assign to the electron equal probability with which it is in each of the four $|\pm_z\rangle$ and $|\pm_y\rangle$ states before the spin measurement time, say $t_m$.

Moreover, the spin state (\ref{eq:superspin}) of the electron can be generally re-written as a superposition of the form,
\begin{equation} 
|\phi\rangle=e^{-i\theta/2}(|+_\theta\rangle + |-_\theta\rangle)  \ ,
\label{eq:superspinth}
\end{equation}
where $|\pm_\theta\rangle$ denote the eigenstates of $\hat{\sigma}_\theta$, the component of the Pauli operator along an arbitrary axis that forms an angle $\theta$ relative to the z-axis in the yz-plane,
\begin{equation}\label{eq:phiexample}
\theta = \left\{ \begin{array}{ll}
\alpha & \ {\mbox{if}} \ \ \ \beta=\pi/2\\
-\alpha & \ {\mbox{if}} \ \ \ \beta=3\pi/2
\end{array} \right. \ ,
\end{equation}
in which $\alpha$ and $\beta$ are the polar and azimuthal angle respectively. Here $|+_\theta\rangle=\cos(\theta/2)|+_z\rangle+i\sin(\theta/2)|-_z\rangle$ and $|-_\theta\rangle=i\sin(\theta/2)|+_z\rangle+\cos(\theta/2)|-_z\rangle$. By invoking {\bf hypothesis 2}, from Eqs. (\ref{eq:superspin}), (\ref{eq:superspiny}), and (\ref{eq:superspinth}) it follows that the electron should be in the spin states $|\pm_z\rangle$, $|\pm_y\rangle$, and $|\pm_\theta\rangle$ states with equal probability unless it undergoes a spin measurement. Following the same rationale leading to Eq. (\ref{eq:superspinyz}), one may obtain a superposition of the following form,
\begin{equation} 
|\phi\rangle=(2\pi)^{-1}\int {\mbox{d}}\theta e^{-i\theta/2} (|+_\theta\rangle + |-_\theta\rangle)   \ ,
\label{eq:superspinthg}
\end{equation}
wherein summation is replaced with the integral over $\theta$ since the angle $\theta$ is a continuous variable. Although Eq. (\ref{eq:superspinthg}) is as odd as Eq. (\ref{eq:superspinyz}), it shows equal probability for the electron to be in each $|\pm_\theta\rangle$ state. Before the spin measurement time $t_m$, according to  Eq. (\ref{eq:superspinthg}), the probability for the electron to be in a specific spin state $|\pm_\theta\rangle$ is $\propto (2\pi)^{-2} {\mbox{d}}^2\theta$, which approaches zero with ${\mbox{d}}^2\theta$. This result may be translated as ``the electron is not in any specific $|\pm_\theta\rangle$ state before the time $t_m$" since the corresponding probability is infinitesimally small, which is understandable since there are infinitely possible spin states for the electron until a $\hat{\sigma}_\theta$-measurement takes place.
Therefore, for a self-consistent interpretation of the superposition principle, the following hypothesis seems reasonable,
\begin{hypothesis}
A particle initially in a state described by Eq. (\ref{eq:superposition}) is not in any $|\varphi_j\rangle$ state until a measurement described by $\hat{A}$ is performed on it at time $t_m$. After the time $t_m$, the particle will be in a final $|\varphi_j\rangle$ state with the probability of $|a_j|^2$. The coefficient $a_j$ quantitatively describes the measurement-induced projection of the initial state into the final one, i.e., $a_j=\langle\varphi_j|\psi\rangle$.
\end{hypothesis}
Although not explicitly stated in the above hypothesis (for the sake of conciseness), the following understanding is implicitly implied: That ``the probability is $|a_j|^2$ with which the particle is in a $|\varphi_j\rangle$ state" is true only before the time, say $t_k$, when the final measurement result becomes known to the experimentalist. After $t_k$, the probability that the particle is in the specific known $|\varphi_j\rangle$ state is 100\% to the experimentalist.

While the general validity of {\bf hypothesis 3} may need further investigation, it suffices for the purpose of the current work. In what follows, let extend the above interpretation of the superposition principle for a single particle to the case of a composite system consisting of two particles $a$ and $b$ in an initial state,
\begin{equation}
|\psi_c\rangle==\sum_{j=1}^m d_j |\varphi_j\rangle_a|\eta_j\rangle_b\ ,
\label{eq:superpositiontwo}
\end{equation}
in which $|\cdot\rangle_{a,b}$ stand for the states of the labeled particles and the state $|\eta_{j}\rangle$ is an eigenstate of an operator $\hat{B}$ describing a measurement on a particle. One should note that the $|\varphi_j\rangle_a$ or $|\eta_j\rangle_b$ state may be degenerate in Eq. (\ref{eq:superpositiontwo}), i.e., $|\varphi_k\rangle_a=|\varphi_l\rangle_a$ or $|\eta_k\rangle_b=|\eta_l\rangle_b$ for $k\ne l$, as long as the basis of $|\varphi_j\rangle_a|\eta_j\rangle_b$ is orthogonal, i.e., ${}_b\langle\eta_k|{}_a\langle\varphi_k|\varphi_l\rangle_a|\eta_l\rangle_b=0$ for $k\ne l$. When $\hat{A}=\hat{B}$, one may have (but not necessarily) $|\varphi_{j}\rangle=|\eta_{j}\rangle$ or $|\varphi_{j}\rangle=|\eta_{m-j+1}\rangle$ in Eq. (\ref{eq:superpositiontwo}). As an illustrative example, let inspect a spin singlet of two electrons,
\begin{equation}
|\phi_c\rangle=(|+_z\rangle_a |-_z\rangle_b - |-_z\rangle_a |+_z\rangle_b)/\sqrt{2}   \ ,
\label{eq:singletz}
\end{equation}
which is a superposed state that can be equivalently re-written as a superposition of the eigenstates of $\hat{\sigma}_{x}$ or $\hat{\sigma}_{y}$,
\begin{eqnarray}
|\phi_c\rangle&=&\ \ (|+_x\rangle_a |-_x\rangle_b - |-_x\rangle_a |+_x\rangle_b)/\sqrt{2}  \nonumber \\
&=&\ \ (|+_y\rangle_a |-_y\rangle_b - |-_y\rangle_a |+_y\rangle_b)/\sqrt{2}\ ,
\label{eq:singletxy}
\end{eqnarray}
or even a superposition of the eigenstates $|\pm_{n}\rangle$ of $\hat{\sigma}_{n}$, the component of the Pauli operator along an arbitrary axis $\hat{n}$ with a polar angle $\alpha$ and an azimuthal angle $\beta$,
\begin{equation}
|\phi_c\rangle=(|+_{n}\rangle_a |-_{n}\rangle_b - |-_{n}\rangle_a |+_{n}\rangle_b)/\sqrt{2}   \ .
\label{eq:singletab}
\end{equation}
Here $|+_{n}\rangle=\cos(\alpha/2)|+_{z}\rangle+\sin(\alpha/2)e^{i\beta}|-_{z}\rangle$ and $|-_{n}\rangle=-\sin(\alpha/2)e^{-i\beta}|+_{z}\rangle+\cos(\alpha/2)|-_{z}\rangle$.

From Eqs. (\ref{eq:singletz}) - (\ref{eq:singletab}), it follows that an observation that ``the composite system initially in a state described by Eq. (\ref{eq:singletz}) is simultaneously in $|+_z\rangle_a |-_z\rangle_b$ and $|-_z\rangle_a |+_z\rangle_b$ states" implicitly implies equal probability with which the composite system is in $|\pm_{x}\rangle_a |\mp_{x}\rangle_b$, $|\pm_{y}\rangle_a |\mp_{y}\rangle_b$, and $|\pm_{n}\rangle_a |\mp_{n}\rangle_b$ states, unless the system is subject to some spin measurement. Because there are infinitely possible spin states for the electrons before the  measurement takes place, the probability for the composite system to be in any specific $|\pm_{n}\rangle_a |\mp_{n}\rangle_b$ state, including the $|\pm_z\rangle_a |\mp_z\rangle_b$ state, must be infinitesimally small. Consequently, one may have a generalized version of {\bf hypothesis 3} for the composite system as follows,
\begin{hypothesis}
A composite system consisting of two particles initially in a state described by Eq. (\ref{eq:superpositiontwo}) is not in any $|\varphi_j\rangle_a|\eta_j\rangle_b$ state until both particles undergo $\hat{A}$-measurement at time $t_{ma}$ and $\hat{B}$-measurement at time $t_{mb}$, respectively. After $t_{ma}$ and $t_{mb}$, the system is in a final state of $|\varphi_j\rangle_a|\eta_j\rangle_b$ with the probability of $|d_j|^2$. The coefficient $d_j$ quantitatively describes the measurement-induced projection of the initial state into the final one, i.e., $d_j=  {}_b\langle\eta_j|{}_a\langle\varphi_j|\psi_c\rangle$.
\end{hypothesis}
With the above hypothesis, one is almost ready to identify the connection between the superposition principle and locality in quantum mechanics. Nevertheless, there is still an important point missing in {\bf hypothesis 4}: What if one particle (electron $a$) is measured while the other (electron $b$) is not? For the composite system of two electrons in the state of (\ref{eq:singletz}), electron $a$ will surely be in either $|+_{z}\rangle_a$ or $|-_{z}\rangle_a$ state with 50\% probability for each state after a $\hat{\sigma}_z$-measurement. As for electron $b$, it has been widely accepted that the particle will be instantaneously projected into a $|-_{z}\rangle_b$ or $|+_{z}\rangle_b$ state according to Eq. (\ref{eq:singletz}) by the $\hat{\sigma}_z$-measurement on electron $a$ and, thereby, the composite system will exhibit a nonlocal nature.

In the following, nonetheless, it will be shown that the observation that ``electron $b$ will be instantaneously projected into a $|-_{z}\rangle_b$ or $|+_{z}\rangle_b$ state according to Eq. (\ref{eq:singletz}) by the $\hat{\sigma}_z$-measurement on electron $a$" lacks solid theoretical proof. To that end, one may expand $|\pm_z\rangle_b$ in terms of $|\pm_x\rangle_b$, $|\pm_y\rangle_b$, or $|\pm_n\rangle_b$, and re-express Eq. (\ref{eq:singletz}) equivalently as
\begin{eqnarray}
|\phi_c\rangle&=&(|+_{z}\rangle_a |+_{x}\rangle_b + |+_{z}\rangle_a |-_{x}\rangle_b)/2 \nonumber \\
&-&(|-_{z}\rangle_a |+_{x}\rangle_b-|-_{z}\rangle_a |-_{x}\rangle_b)/2 \nonumber \\
&=& (-i|+_{z}\rangle_a |+_{y}\rangle_b + |+_{z}\rangle_a |-_{y}\rangle_b)/2 \nonumber \\
&-&(|-_{z}\rangle_a |+_{y}\rangle_b -i|-_{z}\rangle_a |-_{y}\rangle_b)/2 \nonumber \\
&=& \frac{1}{\sqrt{2}}\left(\sin\frac{\alpha}{2}e^{-i\beta}|+_{z}\rangle_a |+_{n}\rangle_b + \cos\frac{\alpha}{2}|+_{z}\rangle_a |-_{n}\rangle_b\right) \nonumber \\
&-& \frac{1}{\sqrt{2}}\left(\cos\frac{\alpha}{2}|-_{z}\rangle_a |+_{n}\rangle_b -\sin\frac{\alpha}{2}e^{i\beta}|-_{z}\rangle_a |-_{n}\rangle_b\right)
\ .\ \ 
\label{eq:singletzxn}
\end{eqnarray}
From the comparison of Eq. (\ref{eq:singletzxn}) with Eq. (\ref{eq:singletz}), it follows that electron $b$ should be instantaneously projected into a $|\pm_{n}\rangle_b$ state with a probability $\propto (1/2)\cos^2(\alpha/2)$ or $(1/2)\sin^2(\alpha/2)$ by the $\hat{\sigma}_z$-measurement on electron $a$, just in the same way as it is reduced into a $|\pm_{z}\rangle_b$ state with a probability $\propto 1/2$. For example, electron $b$ may be instantaneously projected into a $|\pm_{x}\rangle_b$ or $|\pm_{y}\rangle_b$ state with a probability that is half of the probability for electron $b$ to be in a $|\pm_{z}\rangle_b$ state due to the $\hat{\sigma}_z$-measurement on electron $a$. Because there are infinitely possible $|\pm_{n}\rangle_b$ spin states, then the probability is infinitesimally small that electron $b$ is projected into any specific $|\pm_{n}\rangle_b$ spin state, including the $|\pm_{z}\rangle_b$ one, due to the $\hat{\sigma}_z$-measurement on electron $a$. This is in turn contradictory to the previous observation that ``electron $b$ will be instantaneously projected into a $|-_{z}\rangle_b$ or $|+_{z}\rangle_b$ state according to Eq. (\ref{eq:singletz}) by the $\hat{\sigma}_z$-measurement on electron $a$". Accordingly, to be self consistent in theory, one may have the following hypothesis as an important supplement to {\bf hypothesis 4},
\begin{hypothesis}
For a composite system consisting of two particles initially in a state described by Eq. (\ref{eq:superpositiontwo}), particle-{\mbox b} will not be in any $|\eta_j\rangle_b$ state due to an $\hat{A}$-measurement on particle-{\mbox a} at an earlier time $t_{ma}$ until particle-{\mbox b} itself undergoes a $\hat{B}$-measurement later at $t_{mb}>t_{ma}$.
\end{hypothesis}
It seems that {\bf hypothesis 5} comes into light as required by self consistency in theory and reveals a deep connection between the superposition principle and locality. The hypothesis suggests a local nature for the composite system described by Eq. (\ref{eq:superpositiontwo}) and opens the possibility that quantum mechanics may be a local (and statistical) theory. It turns out that violation of Bell inequality cannot serve as a criteria to distinguish locality from nonlocality in quantum mechanics, because the results of relevance in Bell experiments can be well explained by assuming local property for composite system consisting of entangled particles, as will be shown below.

\section{Locality loophole in Bell experiments}
Let consider a Bell experiment involving a composite system of two electrons in a $|\phi_c\rangle$ state given by Eq. (\ref{eq:singletz}). Suppose that the spin of electron $a$ is measured along the z-axis and the spin of electron $b$ is measured along an axis $\hat{n}$ with a polar angle $\alpha$ and an azimuthal angle $\beta$. According to {\bf hypothesis 5}, the $\hat{\sigma}_z$-measurement and $\hat{\sigma}_n$-measurement act locally on electron $a$ and electron $b$, respectively. As a consequence the two particles are independently projected into the $|\pm_{z}\rangle_a$ and $|\pm_{n}\rangle_b$ states. By invoking {\bf hypothesis 4} and Eq. (\ref{eq:singletzxn}), one will have four possible final states for the composite system, $|\pm_{z}\rangle_a |\pm_{n}\rangle_b$ and $|\pm_{z}\rangle_a |\mp_{n}\rangle_b$, with the corresponding probability of $\sin^2(\alpha/2)$ or $\cos^2(\alpha/2)$. Then the expected value of the joint $\hat{\sigma}_z^a\hat{\sigma}_n^b$-measurement on the electrons can be easily calculated as, by use of Eq. (\ref{eq:singletzxn}), 
\begin{eqnarray}
<\hat{\sigma}_z^a\hat{\sigma}_n^b>&=&\langle\psi_c|\hat{\sigma}_z^a\hat{\sigma}_n^b|\psi_c\rangle \nonumber \\
&=& \frac{1}{2}\sin^2\frac{\alpha}{2}{}_a\langle+_z|{}_b\langle+_n|\hat{\sigma}_z^a\hat{\sigma}_n^b|+_z\rangle_a|+_n\rangle_b \nonumber \\
&+&  \frac{1}{2}\cos^2\frac{\alpha}{2}{}_a\langle+_z|{}_b\langle-_n|\hat{\sigma}_z^a\hat{\sigma}_n^b|+_z\rangle_a|-_n\rangle_b \nonumber \\
&+&\frac{1}{2}\cos^2\frac{\alpha}{2}{}_a\langle-_z|{}_b\langle+_n|\hat{\sigma}_z^a\hat{\sigma}_n^b|-_z\rangle_a|+_n\rangle_b \nonumber \\
&+&\frac{1}{2}\sin^2\frac{\alpha}{2}{}_a\langle-_z|{}_b\langle-_n|\hat{\sigma}_z^a\hat{\sigma}_n^b|-_z\rangle_a|-_n\rangle_b  \nonumber\\
&=& \sin^2(\alpha/2)-\cos^2(\alpha/2)   \nonumber\\
&=& - \cos\alpha\ .
\label{eq:singletave}
\end{eqnarray}
Because the choice of coordinate system is arbitrary in the above calculation, Eq. (\ref{eq:singletave}) holds valid for any two axes along which the joint measurement is done. With this result, it is straightforward to prove violation of Bell inequality \cite{Bell1964}: Suppose three arbitrary axes in a plane and the value of the angle between any two of these axes is  respectively $\alpha_{12}$, $\alpha_{23}$, and $\alpha_{13}$. For $\alpha_{12}=45^\circ$, $\alpha_{23}=45^\circ$, and $\alpha_{13}=90^\circ$, By use of Eq. (\ref{eq:singletave}), one has 
\begin{eqnarray} \label{eq:Bellineq}
&& |<\hat{\sigma}_1^a\hat{\sigma}_2^b> - <\hat{\sigma}_1^a\hat{\sigma}_3^b>| -<\hat{\sigma}_2^a\hat{\sigma}_3^b> \nonumber\\
&=& |- \cos\alpha_{12} +  \cos\alpha_{13}| +\cos\alpha_{23} \nonumber\\
&=& |-\cos45^\circ + \cos90^\circ| + \cos45^\circ\nonumber\\
&=&\sqrt{2}>1\ .
\end{eqnarray}

To gain an insightful understanding of the underlying physics behind the resulf of Eq. (\ref{eq:singletave}), let inspect the above Bell experiment from a different point of view. Suppose that one has an electron in a state described by a density operator $\hat{\rho}$,
\begin{equation}
\hat{\rho}=(|+_z\rangle \langle+_z| + |-_z\rangle\langle-_z|)/2 \ .
\label{eq:densityop}
\end{equation}
Now one may perform a {\it thought} experiment in which the electron is subject to a $\hat{\sigma}_z\hat{\sigma}_n$-measurement. The expectation value of the measured quantity reads
\begin{eqnarray}
<\hat{\sigma}_z\hat{\sigma}_n>&=&Tr(\hat{\rho}\hat{\sigma}_z\hat{\sigma}_n)\nonumber \\
&=&2^{-1}(\langle+_z| \hat{\sigma}_z\hat{\sigma}_n |+_z\rangle + \langle-_z| \hat{\sigma}_z\hat{\sigma}_n |-_z\rangle) \nonumber\\
&=& 2^{-1}(\langle+_z| \hat{\sigma}_n |+_z\rangle - \langle-_z| \hat{\sigma}_n |-_z\rangle) \nonumber\\
&=&  \cos\alpha\ ,
\label{eq:spinave}
\end{eqnarray}
wherein $|+_z\rangle=\cos(\alpha/2)|+_n\rangle - \sin(\alpha/2)e^{i\beta}|-_n\rangle$ and $|-_z\rangle=\sin(\alpha/2)e^{-i\beta}|+_n\rangle + \cos(\alpha/2)|-_n\rangle$ are used for the calculation. To successfully implement the {\it thought} experiment, the theoretical result of Eq. (\ref{eq:spinave}) must be mapped to some experimental outcome produced in a measurement scheme with operation steps. The goal may be achieved by the following mathematical manipulations on $<\hat{\sigma}_z\hat{\sigma}_n>$, starting from the second step in the above calculation,
\begin{eqnarray}
<\hat{\sigma}_z\hat{\sigma}_n>
&=& \ \ 2^{-1}\sum_{S_n} \langle+_z| \hat{\sigma}_z |S_n\rangle \langle S_n| \hat{\sigma}_n |+_z\rangle \nonumber\\
&& + 2^{-1}\sum_{S_n} \langle-_z| \hat{\sigma}_z |S_n\rangle \langle S_n| \hat{\sigma}_n |-_z\rangle \nonumber\\
&=& \ \ 2^{-1}(|\langle+_z|+_n\rangle|^2+|\langle-_z|-_n\rangle|^2)
\nonumber\\
&& -2^{-1}(|\langle-_z|+_n\rangle|^2+|\langle+_z|-_n\rangle|^2) \nonumber\\
&=&  2^{-1}\sum_{S_z,S_n} \mbox{sign}(S_zS_n) \ |\langle S_z|S_n\rangle|^2
\ ,
\label{eq:jointmeas}
\end{eqnarray}
in which $\sum_{S_n}|S_n\rangle \langle S_n|=1$ is utilized, $S_n=\pm_n$, $S_z=\pm_z$, $\mbox{sign}(\pm_z\pm_n)=1$, and $\mbox{sign}(\pm_z\mp_n)=-1$. The term $|\langle S_z|S_n\rangle|^2$ in Eq. (\ref{eq:jointmeas}) is the probability that a particle initially in a $|S_n\rangle$ state is projected into a final $|S_z\rangle$ state or vice versa by a $\hat{\sigma}$-measurement, according to {\bf hypothesis 3}. The value of this probability may be obtained in experiment by repeatedly counting the numbers of electron in the $|S_z\rangle$ and $|S_n\rangle$ states, respectively, and then calculating the normalized coincidence detection rate according to
\begin{equation} \label{eq:cc}
|\langle S_z|S_n\rangle|^2= (N_{sz}N_{sn})^{-1/2} \cdot C(N_{sz},N_{sn}) \ ,
\end{equation}
provided that an electron simultaneously in both $|S_z\rangle$ and $|S_n\rangle$ states is available. Here $N_{sz}$ and $N_{sn}$ denote the counted electron numbers in the $|S_z\rangle$ and $|S_n\rangle$ states, respectively, and $C(N_{sz},N_{sn})$ is the corresponding coincidence counting number. When the values of $|\langle S_z|S_n\rangle|^2$ obtained using Eq. (\ref{eq:cc}) are plugged into Eq. (\ref{eq:jointmeas}), one will have the measured value of $<\hat{\sigma}_z\hat{\sigma}_n>$ as predicted by Eq. (\ref{eq:spinave}).

\begin{figure}[htbp]
\centering
\includegraphics[width=8cm]{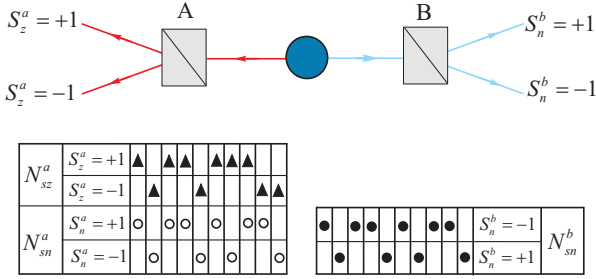}
\caption{(color online)  Illustration of how a local $\hat{\sigma}_z\hat{\sigma}_n$-measurement on an electron (electron $a$) may be realized by use of quantum entanglement. The electrons in a state given by Eq. (\ref{eq:singletz}) are separated and fly in opposite directions. The spin of electron $a$ ($b$) is measured independently at A (B) along the z-axis ($\hat{n}$-axis), producing electron number counting results for $N_{sz}^a$ ($N_{sn}^b$) as indicated by black triangles (black dots). Because the number counting data for $N_{sn}^a$ (circles) are unavailable, one may use the number counting data $N_{sn}^b$ (black dots) as a duplicate of the unavailable $N_{sn}^a$ data in the production of the result of the local $\hat{\sigma}_z\hat{\sigma}_n$-measurement on electron $a$ using Eqs. (\ref{eq:ccc}) and (\ref{eq:jointmeasa}).}
\label{fig:cc}
\end{figure}

Unfortunately, no electron simultaneously in both the $|S_z\rangle$ and $|S_n\rangle$ states is available in the above {\it thought} experiment; after all, it cannot be detected in more than one state at a time. To overcome this problem one may resort to the composite system of two electrons described by Eq. (\ref{eq:singletz}). To be specific, let assume that a $\hat{\sigma}_z^a\hat{\sigma}_n^a$-measurement is to be performed on electron $a$, which is in a state described by a reduced density operator as given by Eq. (\ref{eq:densityop}). First, let electrons in the same initial state as electron $a$ be repeatedly sent to the particle counters that output the numbers of electron  $N_{sz}^a$ in the final $|\pm_z\rangle_a$ states. At the same time, electrons in the same initial state as electron $b$ are sent to other particle counters for the numbers of electron in the final $|\pm_n\rangle_b$ states. Although measurements on the two electrons are carried out locally (and independently), due to the quantum correlation between the two electrons determined by Eq. (\ref{eq:singletab}), the number counting result $N_{sn}^b$ for $|\mp_n\rangle_b$ electrons (electron $b$) can be used as a replica of the number counting result $N_{sn}^a$ for $|\pm_n\rangle_a$ electrons (electron $a$), i.e., $N_{\pm n}^b=N_{\mp n}^a$, the latter of which is practically unavailable in the experiment (Fig. \ref{fig:cc}). Therefore, in the calculation of the coincidence detection rate for electron $a$ using Eq. (\ref{eq:cc}), one will have
\begin{eqnarray}\label{eq:ccc}
|\langle S_z^a|-S_n^a\rangle|^2&=&|\langle S_z^a|S_n^b\rangle|^2 \nonumber\\
&& = (N_{sz}^a\ N_{sn}^b)^{-1/2} \cdot C(N_{sz}^a,N_{sn}^b) \ .
\end{eqnarray} 
From Eqs. (\ref{eq:jointmeas}) and (\ref{eq:ccc}), it follows that 
\begin{eqnarray}
<\hat{\sigma}_z^a\hat{\sigma}_n^a>
&=&  2^{-1}\sum_{S_z^a,S_n^a} \mbox{sign}(S_z^a S_n^a) \ |\langle S_z^a|S_n^a\rangle|^2\nonumber\\
&=&  -2^{-1}\sum_{S_z^a,S_n^b} \mbox{sign}(S_z^a S_n^b) \ |\langle S_z^a|S_n^b\rangle|^2\nonumber\\
&=&-<\hat{\sigma}_z^a\hat{\sigma}_n^b> \ .
\label{eq:jointmeasa}
\end{eqnarray}
One should note that the minus sign of $<\hat{\sigma}_z^a\hat{\sigma}_n^b>$ in comparison with $<\hat{\sigma}_z^a\hat{\sigma}_n^a>$ comes from the fact that the spin of electron $a$ is opposite to that of electron $b$, i.e., $S_n^a=-S_n^b$. 

Eq. (\ref{eq:jointmeasa}) shows that a joint $\hat{\sigma}_z^a\hat{\sigma}_n^b$-measurement on the two electrons described by Eq. (\ref{eq:singletz}) provides an elegant way to realize a local $\hat{\sigma}_z^a\hat{\sigma}_n^a$-measurement on electron $a$ when the minus sign is neglected, which is otherwise a task difficult to accomplish. Both results of joint $\hat{\sigma}_z^a\hat{\sigma}_n^b$-measurement and local $\hat{\sigma}_z^a\hat{\sigma}_n^a$-measurement violate Bell inequality, which explains the physics behind the result of Eq. (\ref{eq:singletave}) and proves the existence of locality loophole in Bell experiments. In view of the above analysis, a concept of physical nonlocality is suggested and may be referred to as ``state collapse of one particle due to a measurement on the other at a space-like distance when the two particles are in a state given by Eq. (\ref{eq:superpositiontwo})"; meanwhile, a concept of mathematical nonlocality is suggested and may be referred to as ``violation of Bell inequality". With these two concepts, it is appropriate to state that Bell experiments have successfully demonstrated mathematical nonlocality, while the experimental results of relevance may be well explained with the concept of physical locality.

At this point, one may argue that the locality loophole may be closed by invoking the uncertainty principle as follows. Physical locality will allow one to simultaneously perform a $\hat{\sigma}_z^a$-measurement on electron $a$ and a $\hat{\sigma}_y^b$-measurement on electron $b$. Since the two measurements are carried out locally and independently, one may have arbitrarily precise results for both measurements, $\hat{\sigma}_z^a$ and $\hat{\sigma}_y^b$. Given the quantum correlation as shown by Eq. (\ref{eq:singletxy}), the value of $\hat{\sigma}_y^a$ may be precisely inferred from the $\hat{\sigma}_y^b$ measurement, leading to precise measurement of both $\hat{\sigma}_z^a$ and $\hat{\sigma}_y^a$ and hence violating the uncertainty principle \cite{epr}. Nonetheless, the above rationale has overlooked the result of Eq. (\ref{eq:singletave}) from which it follows that the joint $\hat{\sigma}_z^a\hat{\sigma}_y^b$-measurement always yields an expectation value of zero; this value can be seriously contaminated by any noise at a finitely small level in practice leading to very poor signal-to-noise ratio for the measurement. Alternatively, one may gain an understanding of the problem by noting that, whenever electron $a$ is detected in a local $\hat{\sigma}_z^a$-measurement at an earlier time $t_{m}$, electron $b$ will be randomly projected into $|\pm_y\rangle$ states with equal probability of 50\% in a later local  $\hat{\sigma}_y^b$-measurement after time $t_{m}$, according to Eq. (\ref{eq:singletzxn}). In other words, the local $\hat{\sigma}_y^b$-measurement on electron $b$ will not gain useful spin information practically after the earlier local $\hat{\sigma}_z^a$-measurement on electron $a$ and, hence, violation of the uncertainty principle is avoided.

\section{Existing experimental results that indicate locality}
The loophole of physical locality in Bell experiments, if not successfully closed, may have far-reaching impacts on our understanding of quantum mechanics. To consolidate the nonlocality concept, one needs experimental evidences beyond what has been demonstrated by Bell experiments. What happened sounds opposite to the above expectation and makes the situation more serious: There exist reported experimental results \cite{menzel2012,menzel2013} that seems to favor physical locality in quantum mechanics; the results indicate that a measurement on one of two entangled photons caused no state collapse to its partner. In what follows, let briefly present the experimental results reported ten years ago and then discuss how these results are related to locality.

\begin{figure}[htbp]
\centering
\includegraphics[width=8.5cm]{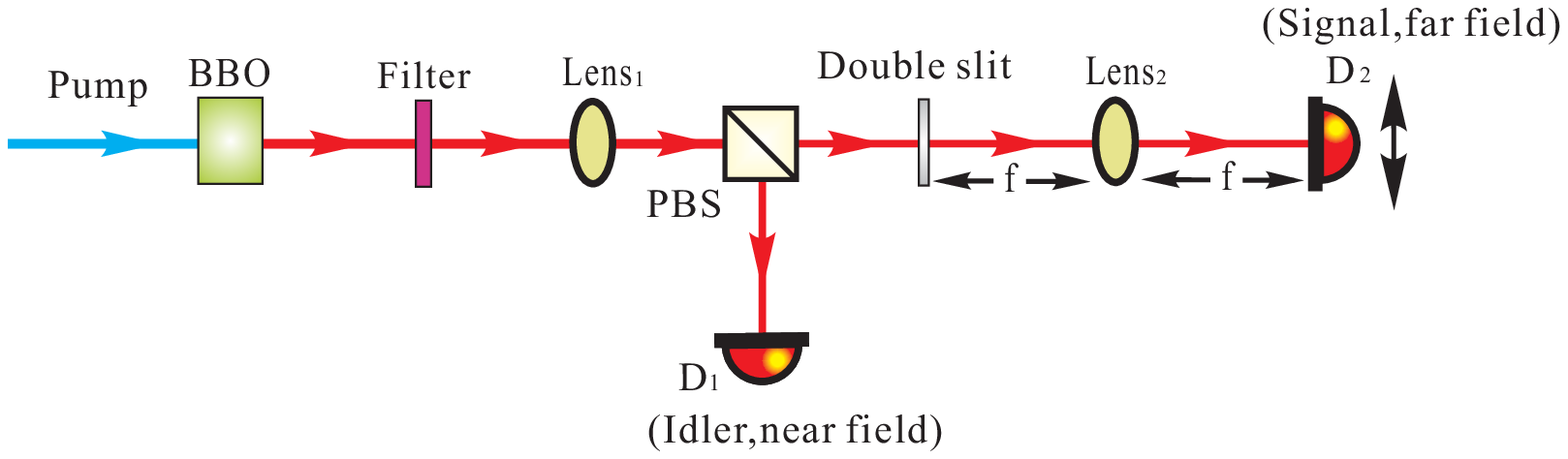}
\caption{(color online)  Experimental schematics originally on simultaneous observation of which-slit information and interference with entangled photons \cite{menzel2013}. Entangled photons were generated through a SPDC process in a beta barium borate (BBO) crystal with a TEM$_{01}$ mode pump. The photons passed through a narrow band filter and a lens (lens1) imaged the two intensity maxima at the exit of the crystal onto the openings of the double slit or detector D$_1$. The far-field detection of photons (photon $s$) is realized by a lens (lens2) in an f-f arrangement between the double slit and detector D$_2$, the later of which is spatially scanned. PBS: Polarizing beamsplitter.}
\label{fig:spdc}
\end{figure}

In a double-slit configuration, two experiments were conducted in a row on wave-particle dualism and complementarity by use of entangled photons \cite{menzel2012,menzel2013}. The photon pairs were produced through Type-II spontaneous parametric down-conversion (SPDC) as depicted in Fig. \ref{fig:spdc}. The unique feature shared by the experiments was the use of laser pumps in TEM$_{01}$ modes, each displaying two distinct intensity maxima in an upper-lower geometry. As a consequence, the signal and idler beams with orthogonal polarization were created in TEM$_{01}$ modes as well. In each experiment, since the average intensity of the pump laser was controlled low enough, only one pair of photons was created during each measurement time interval. A lens was used to couple the generated photons into the openings of the double slit (signal, photon $s$) or directly into a photon detector D$_1$ (idler, photon $i$).

Under the above experimental conditions, the created photon pairs  as a composite system in a near-field plane were in an entangled state,
\begin{equation}
|\phi_c\rangle=(|u\rangle_s |u\rangle_i + e^{i\gamma} |l\rangle_s |l\rangle_i)/\sqrt{2}   \ ,
\label{eq:entphoton}
\end{equation}
in which $|u\rangle$ ($|l\rangle$) denotes the state where a photon is in the upper (lower) intensity maxima of the TEM$_{01}$ mode and $\gamma$ is an arbitrary phase angle determined by the SPDC process. The quantum correlation delineated by Eq. (\ref{eq:entphoton}) originating from the fact that the paired photons were simultaneously generated at the same spatial point was experimentally confirmed with 99$\%$ fidelity \cite{menzel2012}. From the concept of physical nonlocality and the observed correlation in experiment, it has been widely believed that the detection of one photon (photon $i$) in a lower intensity maximum of the TEM$_{01}$ mode would cause a state collapse to the other (photon $s$) that should be also in the lower intensity maximum \cite{menzel2012,menzel2013}. Because of this measurement-induced state reduction, photon $s$ should lose its coherence between the two intensity maxima, preventing one from observing interference with known which-slit information. However, what was observed in the experiments is very interesting: The authors moved another detector, D$_2$, into the far field of the double slit and measured photon $s$ as a function of its vertical position in coincidence with photon $i$ recorded by the detector D$_1$ placed at the lower intensity maximum of the TEM$_{01}$ mode \cite{menzel2012}. They observed interference fringes of photon $s$ with a visibility of about 50\%, which was much higher than the expected value of 1\% for the fringe visibility given the 99\% which-slit information provided by detector D$_1$. 

To understand the above experimental results, one must note that they  are repeatable and not contradictory to other experiments on similar topic \cite{kaiser2012,Bienfait2020}. Usually, detection of the photons should be implemented in a manner such that their quantum correlation remains as they are produced, which is exactly what was done in \cite{kaiser2012,Bienfait2020}. If so, after both were detected, the photon pairs would be projected into a $|u\rangle_s |u\rangle_i$ or $|l\rangle_s |l\rangle_i$ state with equal probability, according to {\bf hypothesis 4} and Eq. (\ref{eq:entphoton}). Indeed, this correlation was observed as reported in \cite{menzel2012,menzel2013} when both photons were collected in near field. What distinguishes the presented experiments from others is that the authors detected photon $i$ in near field while photon $s$ was collected in the far field of the double slit. As a consequence, the quantum correlation between the paired photons given by Eq. (\ref{eq:entphoton}) were lost in the measurement since this correlation remains only in near field.

Now the above experimental results can be explained using the concept of physical locality developed in this work. From {\bf hypothesis 5} it immediately follows that, despite the quantum correlation of the photon pairs given by Eq. (\ref{eq:entphoton}), the local detection of photon $i$ in a $|u\rangle_i$ or $|l\rangle_i$ state (in a upper or lower maximum of a TEM$_{01}$ mode) did not cause the usually expected state collapse to photon $s$ at the double slit in near field where photon $s$ was not in a $|u\rangle_s$ or $|l\rangle_i$ state. Instead, it passed through the double slit in a state as it was created and arrived in the far-field area where it was not actually detected; therefore, photon $s$ kept its own coherence and power balance between the intensity maxima at the double slit, enabling observation of its interference fringes in far field as reported in \cite{menzel2012,menzel2013}.

The above experimental results were attributed by the authors to their special choice of the TEM$_{01}$ mode pumps: The mode function, including the “unoccupied” intensity maximum, was imprinted on photon $s$ in form of a superposition of two wave vectors giving rise to the observed interference fringes \cite{menzel2012}. However, the said mode function is only one of the necessary, but not sufficient, conditions for the interference fringes to appear. Other necessary conditions include at least coherence and power balance between the two humps of the TEM$_{01}$ mode. From the results of the observed  interference fringes, it follows that both intensity maxima of the TEM$_{01}$ mode for the signal beam must have been occupied with approximately balanced light power, i.e., there was no “unoccupied” intensity maximum in the signal beam. This indicates that no state reduction happened to photon $s$ after its partner was detected; otherwise its coherence and power balance should have been lost between the two intensity maxima. The detection of photon $i$ by detector D$_1$ provided no which-slit information for photon $s$ and, therefore, the experiments reported in \cite{menzel2012,menzel2013} seem to be of little relevance to wave-particle dualism and complementarity; instead, they might be the first experimental results that indicate physical locality in quantum mechanics: The detection of the idler photon caused no state collapse to the unmeasured signal photon despite of their quantum correlation.

\section{Proposed experiments}
\begin{figure}[htbp]
\centering
\includegraphics[width=8.5cm]{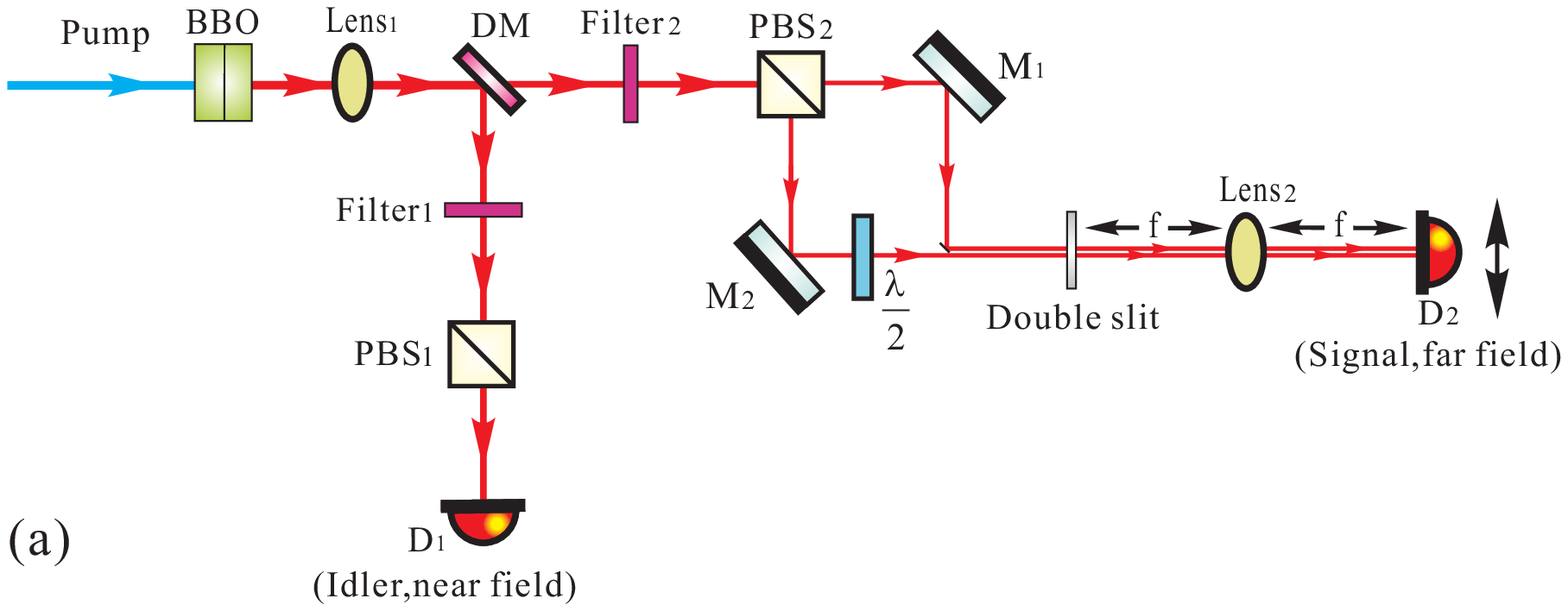} \\
\vspace{0.2in}
\includegraphics[width=8.5cm]{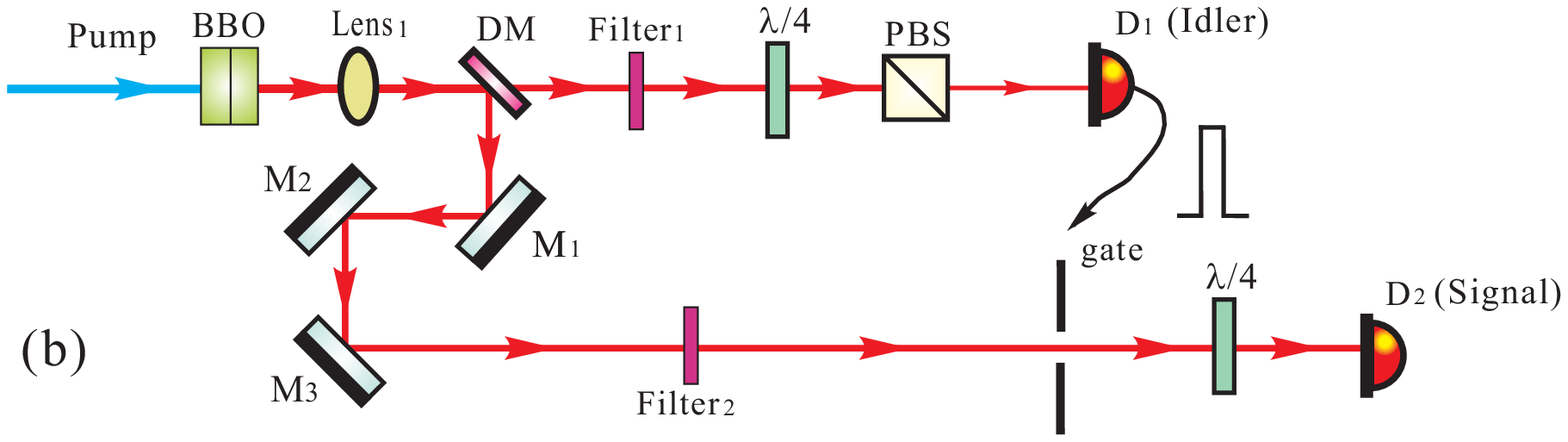}
\caption{(color online) Proposed experiments on physical locality with polarization entangled photons. Photon pairs are generated through type-I SPDC with two BBO crystals. One of the two crystals are rotated relative to the other by 90$^\circ$ around the beam propagation direction. Each photon pair is separated by a dichromatic mirror (DM) and the separated photons then pass through two narrow band filters respectively. (a) The polarization of the idler photon is measured in near field, whereas spatially scanning far-field detection of photon $s$ is realized by a lens (lens2) in an f-f arrangement between the slits and detector D$_2$. M$_{1,2}$: Reflective mirror. (b) After the circular polarization of the idler photon is measured, a pulse signal is sent to open a gate so that the signal photon may pass through a quarter-wave plate on which the photon in a CP state will exert a mechanical torque.  M$_{1-3}$: Reflective mirror. $\lambda/4$: Quarter-wave plate.}
\label{fig:prop1}
\end{figure}

Given the above experimental results and the fundamental importance of the nonlocality concept in quantum mechanics, more theoretical and experimental investigations are highly desired to close the loophole of physical locality. To that end, two kinds of new experiment are proposed as follows. The first type of experiment (Fig. \ref{fig:prop1}a) makes use of paired photons, $s$ and $i$, with different wavelengths in a polarization entangled state,
\begin{equation}
|\phi_c\rangle=(|H\rangle_s |H\rangle_i + |V\rangle_s |V\rangle_i)/\sqrt{2}   \ ,
\label{eq:polent}
\end{equation}
where $|H\rangle$ and $|V\rangle$ stand for the horizontal and vertical polarization states respectively. The entangled photons are separated by a dichromatic mirror and the polarization of photon $i$ is measured in near field. The signal beam carrying photon $s$ is split into two parts by a polarizing beamsplitter and the polarization of one part of the signal beam is rotated by $90^\circ$ with a half-wave plate. Then the two parts of the signal beam are combined before a double slit after which the combined beam propagates to the far field where a scanning detector is used to collect the signal photon as a function of the scanning position. The key idea of this experiment is to avoid a special choice of TEM$_{01}$ mode pump and study how the polarization measurement on photon $i$ in the near field may affect the interference fringes of photon $s$ in the far field. Observation of interference will verify physical locality, which will otherwise be rejected if no expected interference shows up (locality loophole closed).

The second type of experiment (Fig. \ref{fig:prop1}b) exploits paired photons in the same state given by Eq. (\ref{eq:polent}) which can also be re-written as,
\begin{equation}
|\phi_c\rangle=(|R\rangle_s |L\rangle_i + |L\rangle_s |R\rangle_i)/\sqrt{2}   \ ,
\label{eq:cpent}
\end{equation}
where $|L\rangle=(|H\rangle-i|V\rangle)/\sqrt{2}$ and $|R\rangle=(|H\rangle+i|V\rangle)/\sqrt{2}$ denote the left-hand and right-hand circular polarization (CP) states, respectively. Again the entangled photons have different wavelengths and are separated by a dichromatic mirror. The essence of this experiment is to explore the mechanical detection of photon's angular momentum if the light beam is in a  CP state \cite{Beth1936}. When the composite system owns nonlocal property, detection of one photon (idler) in a CP state, according to Eq. (\ref{eq:cpent}), will immediately project its partner (signal) into another CP state. Since a circularly polarized photon carries nonzero angular momentum, it will exert a mechanical torque on a half-wave plate when passing through it and, thereby, the CP state of the signal photon can be mechanically measured. Therefore, through observation of this torque effect on the half-wave plate exerted by the signal photon, one may determine whether photon $s$ is projected into an expected CP state by the detection of photon $i$ in another CP state as dictated by Eq. (\ref{eq:cpent}). This experiment will provide more information that may be very useful to close the loophole of physical locality. 




\section{Conclusions}
As part of the efforts to consolidate the nonlocality concept in quantum mechanics, this work has revealed a connection between the superposition principle and locality. A self consistent interpretation of the superposition principle has been put forth, from which it has been shown that quantum mechanics may be a local statistical theory in principle. It has been shown that Bell experiments can be satisfactorily explained by assuming local nature for entangled particles, which means that Bell inequality cannot be used as a convincing method to distinguish locality from nonlocality and is referred to as the loophole of physical locality. More seriously, existing experimental results have been presented in favor of locality in quantum mechanics, which urges both theoretical and experimental investigations to close the loophole. To that end, two types of experiment have been proposed.

\begin{acknowledgments}
This work was financially supported by the National Natural Science Foundation of China (grant number 12074110). The author is grateful to Mr. Wu for preparing all the figures.
\end{acknowledgments}

\nocite{*}

\bibliography{apssamp}

\end{document}